\documentclass[twoside,12pt]{article}
\usepackage{epsfig}

\def\Journal#1#2#3#4{{#1} {#2} (#4) #3 }

\def\NPB{{\em Nucl. Phys.} B}

\def\PLB{{\em Phys. Lett.} B}

\def\PRL{\em Phys. Rev. Lett.}
\def\PREV{\em Phys. Rev.}

\def\PRD{{\em Phys. Rev.} D}

\newcommand{\be}{\begin{equation}}
\newcommand{\ee}{\end{equation}}
\newcommand{\bea}{\begin{eqnarray}}
\newcommand{\eea}{\end{eqnarray}}

\newcommand{\hbo}{\hbox to 1 true cm {\hfill } } 
\newcommand{\tr}{\mathrm{tr} \, } 
\newcommand{\Det}{\mathrm{Det} \, } 
\def\dslash{\partial\kern-.5em\slash}
\def\I{{\rm i}}
\def\E{\mathrm{e}}

\begin{document}

\title{ \vspace{1cm} From confinement to new states of dense QCD matter }
\author{Kurt Langfeld$^{\ 1}$, Andreas Wipf$^{\ 2}$ \\
\\
$^1$ School of Computing \& Mathematics, University of Plymouth, 
Plymouth, UK \\
$^2$Theoretisch-Physikalisches Institut, Friedrich-Schiller-Universit\"at, 
Jena, Germany } 
\maketitle
\begin{abstract} Transitions between centre sectors are related to 
confinement in pure Yang-Mills theories. We study the impact of these 
transitions in QCD-like theories for which centre symmetry is {\it explicitly} 
broken by the presence of matter. For low temperatures, we provide numerical
evidence that centre transitions do occur with matter merely providing a 
bias towards the trivial centre sector until centre symmetry is {\it
spontaneously } broken at high temperatures. The phenomenological consequences
of these transitions for dense hadron matter are illustrated in the Schwinger
model and an SU(3) effective quark theory: centre dressed quarks undergo
condensation due to Bose-type statistics forming a hitherto unknown state of
dense but cold quark matter. 
\end{abstract}
\section{Introduction}
Pure Yang-Mills theories feature a colour-confinement phase at low
temperatures separated by a phase transition from the gluon plasma phase at 
high temperatures. This phase structure is well understood by virtue of the 
centre symmetry which is realised at low but spontaneously broken at high
temperatures~\cite{Svetitsky:1982gs,Svetitsky:1985ye}. The (temporal) Polyakov
line transforms homogeneously under a centre transformation of the gluon
fields, and, thus, its expectation value serves as an order parameter for
confinement.  

\vskip 0.3cm 
The situation is less clear cut in QCD-like theories where 
matter transforming under the {\it fundamental} representation of 
the gauge group is coupled to the $SU(N_c)$ gluon fields. Here, 
centre symmetry is explicitly broken and, strictly speaking, 
quark confinement is absent since the colour-electric flux tube 
between a static quark antiquark pair can break by means of the creation 
of a dynamical (light) quark antiquark pair. The string breaking distance 
$r_B$ can be estimated by 
$$
r_B \;\approx \; 2 \, M  \, / \, \sigma \; , 
$$
where $M$ is the (constituent) mass of the dynamical matter, and 
$\sigma $ is string tension of the corresponding pure Yang-Mills theory. 
Thus, in the case for heavy matter, QCD-like theories at intermediate 
scales, i.e., $1/\sqrt{\sigma } < r< r_B$, are well described by their 
pure Yang-Mills counter part. The Polyakov line is almost disordered by 
the approximate centre symmetry, and its expectation value is only 
non-zero due to bias towards the trivial centre sector by the matter 
fields. 

\vskip 0.3cm 
In certain gauges and using lattice gauge simulations, confinement 
has been successfully attributed to certain gluonic degrees 
of freedom. For later use, we would like to mention the centre vortex picture 
(see~\cite{Greensite:2003bk} for a review): vortices experience 
an intrinsically fine-tuned balance between energy and 
entropy~\cite{Zakharov:2006vt} implying that their intrinsic energy scale 
is set by the physical confinement scale~\cite{Langfeld:1997jx}
(rather than the UV regulator of the theory). It was argued recently
that vortex configurations are the image (in the maximal centre gauge) of 
smooth instanton-like configurations (which do confine) thus explaining 
the intrinsic fine-tuning between vortex energy and
entropy~\cite{Langfeld:2010nm}. Vortex percolation is directly 
related to the disorder of the centre symmetry, and vortex de-percolation 
explains the finite temperature deconfinement transition to the hot gluon
plasma phase~\cite{Langfeld:1998cz,Engelhardt:1999fd,Langfeld:2003zi}. 

\vskip 0.3cm 
The approximate centre symmetry in QCD-like theories has strong 
phenomenological consequences if dense matter is considered: 
At least for theories with an even number of colours, quarks 
which are subjected to certain gluonic background fields acquire 
Bose statistic and can undergo condensation~\cite{Langfeld:2009cd}, 
the so-called {\it Fermi-Einstein condensation }~\cite{Langfeldect2010}. 
Quite recently, this mechanism has be thoroughly studied using 
the exactly solvable Schwinger model and resorting to effective quark 
models. It was found~\cite{Langfeld:2011rh} that the mechanism extends its
scope to  theories with an odd number of colours (though pressure needs to  
be applied to the dense medium) and it was conjectured that 
a new state of cold, but dense matter might exist in the hadronic phase 
of QCD for which Fermi Einstein condensation is realised.

\vskip 0.3cm 
In this paper, we review the properties  of the 
approximate centre symmetry and the line of arguments leading 
to Fermi-Einstein condensation. We review the exact result from 
th Schwinger-model in~\cite{Langfeld:2011rh} and discuss the 
possibilities to turn the SU(3) quark model of~\cite{Langfeld:2011rh} 
from minimal to realistic.

\section{Fermi-Einstein condensation }

\subsection{ \it Yang-Mills moduli and vacuum structure of QCD-like theories}

\begin{figure}[tb]
\begin{center}
\begin{minipage}[t]{8 cm} 
\centerline{ 
\epsfig{file=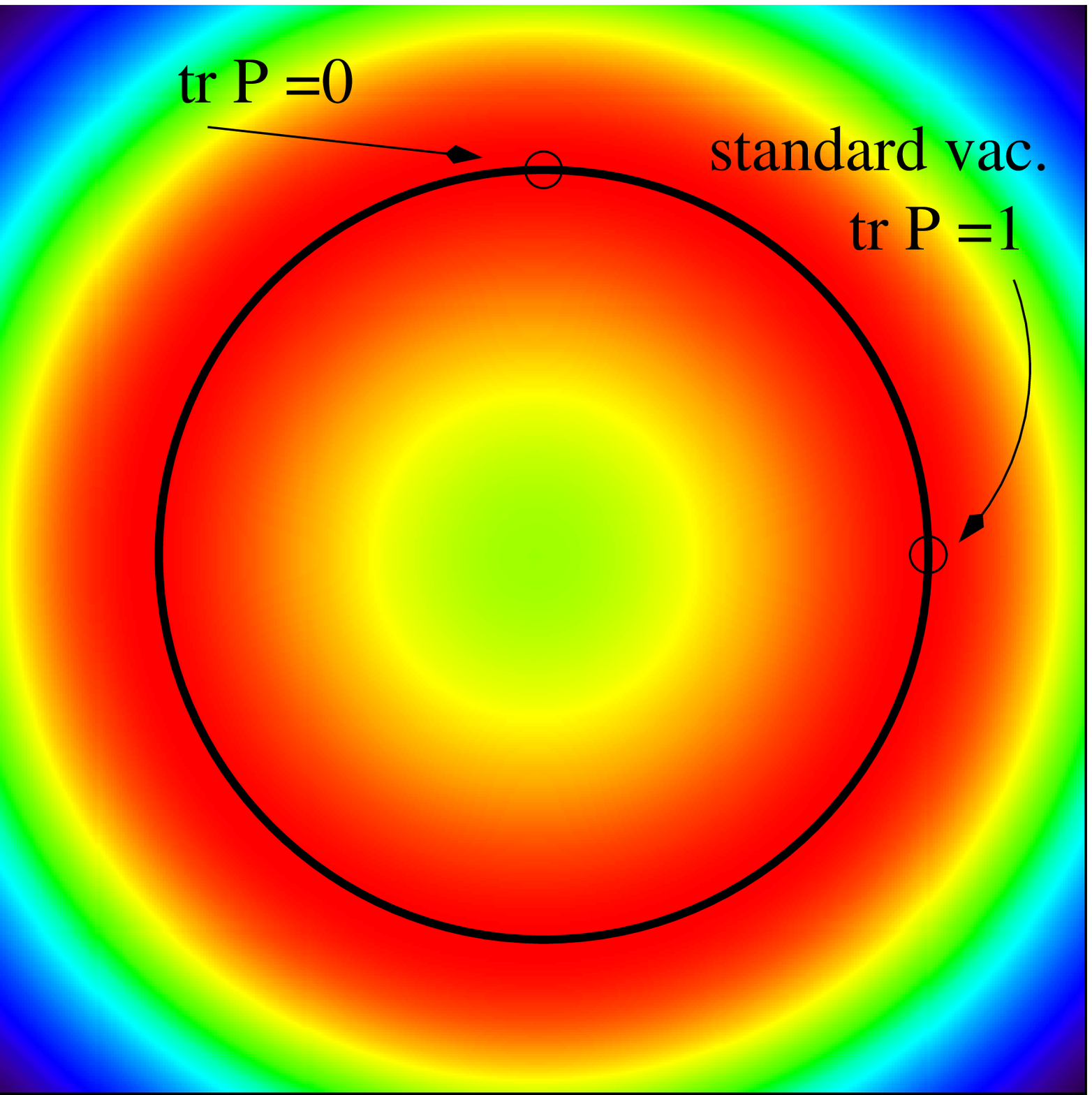,scale=0.3}
\epsfig{file=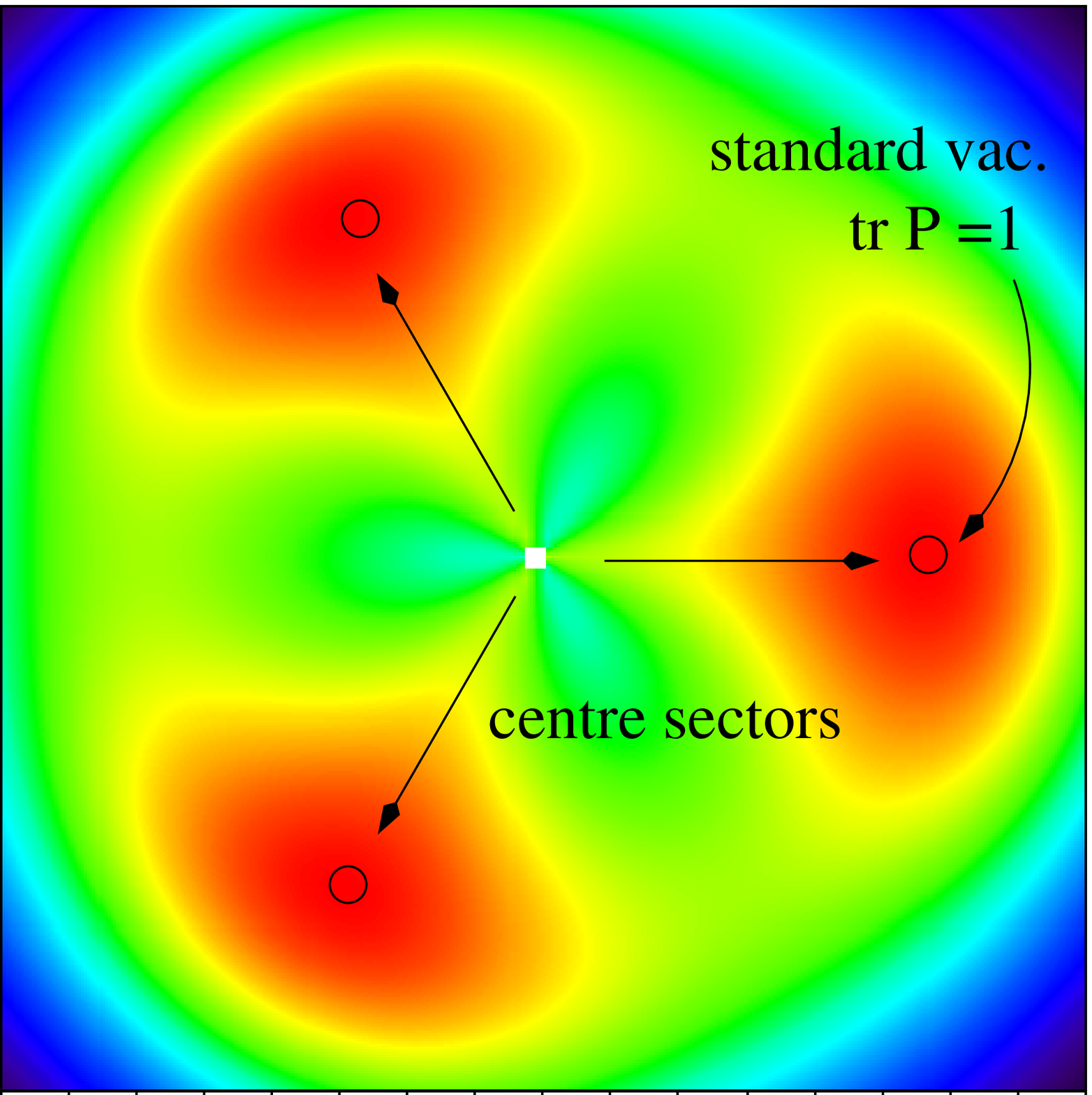,scale=0.3}
\epsfig{file=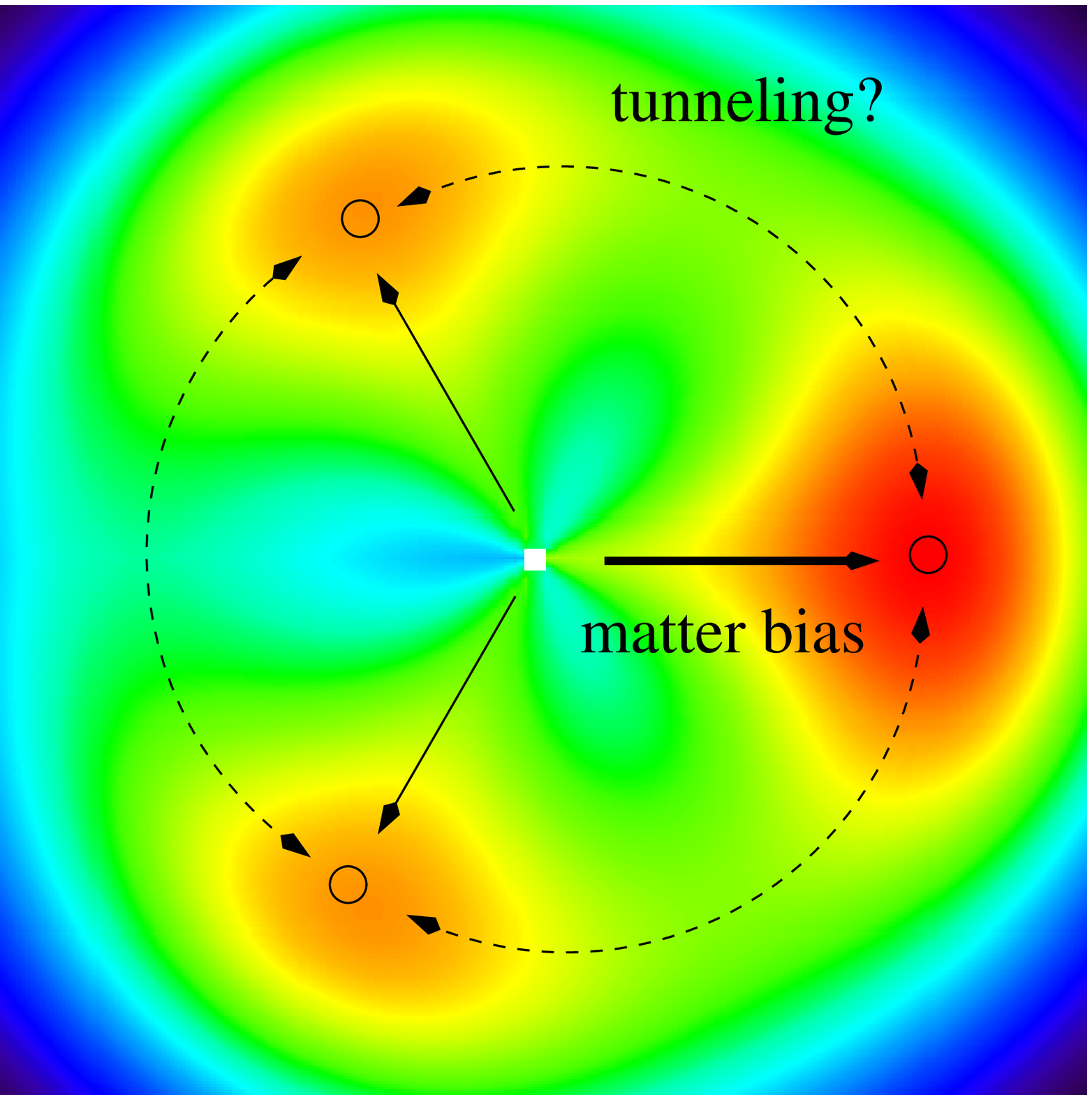,scale=0.3}}
\end{minipage}
\begin{minipage}[t]{16.5 cm}
\caption{Sketch of the {\it classical action} as function of the 
 Yang-Mills configuration space (left); {\it effective action } of pure 
 Yang-Mills theory (middle); {\it effective action } of QCD-like theories,  
 i.e., YM theory with dynamical matter, (right).\label{fig:1}}
\end{minipage}
\end{center}
\end{figure}
Let us firstly discuss pure Yang-Mills theory on a toroidal space-time. 
We will call an {\it empty vacuum} a gauge field configuration $A_\mu (x)$ for
which  any holonomy along a contractible loop ${\cal C}$ is the unit element: 
\be 
W_{\cal C} \; = \;  {\cal P} \; \exp \left\{ i \int _{\cal C} A_\mu (x) \; dx_\mu
\right\} \; = \;  1 , \hbo {\cal P}: \; \hbox{path ordering.} 
\label{eq:1}
\ee 
How many of these configurations are there? Obviously, $A_\mu (x) =0$ 
is an example, and since $W_{\cal C} $ is an element of the adjoint 
representation, all gauge equivalents of this configuration are also 
empty vacua. The more interesting question is then: how many 
{\it gauge inequivalent} field configurations are there which all satisfy 
(\ref{eq:1}). It turns
out~\cite{Keurentjes:1998uu,Selivanov:1999cr,Schaden:2004ah} 
that at least a smooth $U(1)^{4(N_c-1)}$  manifold of gauge inequivalent 
configurations exists (see e.g.~\cite{Langfeld:2011rh}). For a map 
of this manifold, we define the Polyakov line in $\mu $-direction by 
\be 
P_{(\mu)}(x) \; = \;  {\cal P} \; \exp \left\{ i \int _{\ell} A_\mu (x) \; dx_\mu
\right\}   \hbo \hbox{(no sum over $\mu $), } 
\label{eq:2}
\ee 
where $\ell $ is the straight line starting at $x$ winding through 
the torus in $\mu $ direction. The empty-vacuum configurations are then 
labeled by constant Polyakov lines which take values in the Cartan subgroup 
of the gauge group. 
$$ 
P_{(\mu)}(x) \; = \; P_{(\mu)} \; \in \; \hbox{Cartan subgroup} , 
\hbo \mu =1 \ldots 4 \; . 
$$
The situation is illustrated in figure~\ref{fig:1}, left panel, which 
is a cartoon of the classical action as a function of the 
gauge invariant configuration space (all gauge equivalent configurations 
are identified with one point of the canvas). The manifold of least action 
is indicated by the circle. Also indicated is the standard perturbative 
vacuum $A_\mu (x)=0$ as well as an empty vacuum configuration 
with a vanishing trace of the Polyakov line. One can show (see 
e.g.~\cite{Langfeld:2011rh}) that the Polyakov line correlator, 
calculated with any of the empty vacuum configurations ${\cal E}$, 
$$ 
\Bigl[ \tr P_{(\mu)}(x) \; \tr P_{(\mu)}(y) \Bigr] _{\cal E} 
\; = \; \hbox{constant} \; , 
$$
and, hence, does not support a non-trivial quark antiquark potential. 
Below, however, we will argue that a patch configuration which consists 
of regions in space-time with different empty-vacuum-configurations, 
is well suited to describe confinement. 

\vskip 0.3cm 
\begin{figure}[tb]
\begin{center}
\begin{minipage}[t]{8 cm} 
\centerline{ 
\epsfig{file=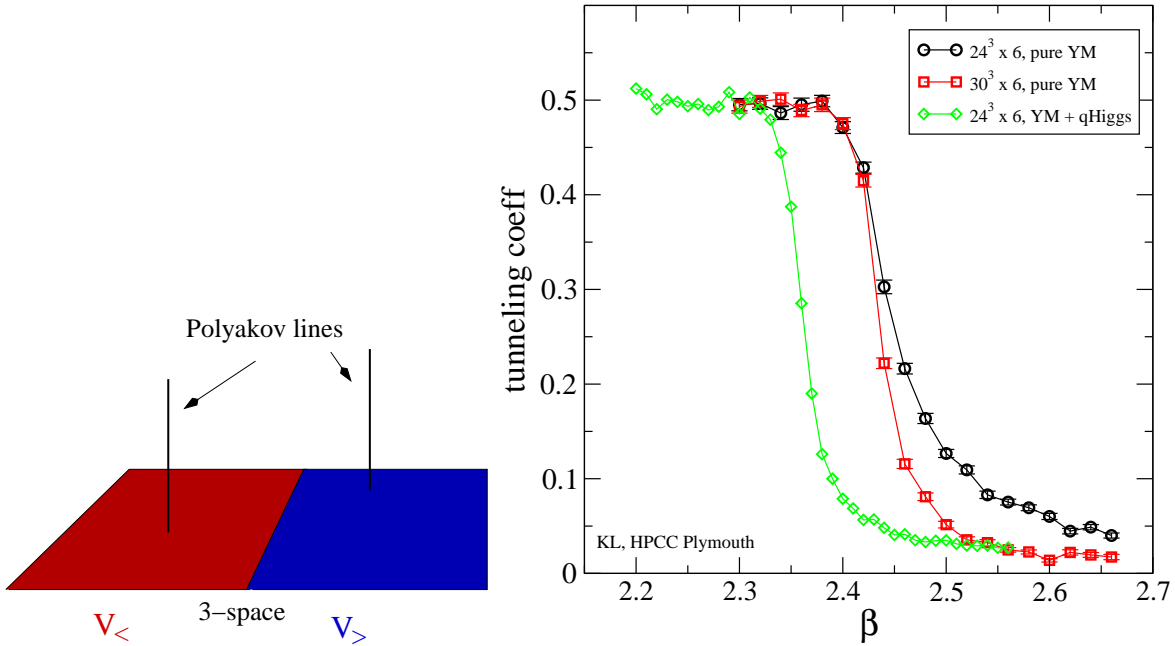,scale=0.4}
\epsfig{file=tunn.eps,scale=0.5}
}
\end{minipage}
\begin{minipage}[t]{16.5 cm}
\caption{Left: Illustration of the ``Litmus paper'' for centre sector 
transitions (left). Right: The tunneling coefficient  $p$ as a function of
the Wilson  $\beta $ for the pure SU(2) gauge theory (black and red) and for
the SU(2) Higgs theory (plot from~\cite{Langfeld:2011rh}).
\label{fig:2}}
\end{minipage}
\end{center}
\end{figure}
At quantum level, the Coleman effective action as a function 
of the so-called classical gluon field needs to be considered, and the 
effective action significantly differs from the classical action 
for a strongly coupled theory. For the pure SU(3) Yang-Mills theory, 
such an effective action is sketched in figure~\ref{fig:1} (picture in the
middle).  The ``flat directions'' of the classical action are lifted 
by the quantum fluctuations. Note, however, that the discrete $Z_3$ centre
transformations ($N_c=3$ below), e.g., 
\be 
A_\mu (x) \to A_\mu ^z (x) \; = \; A_\mu (x) + \frac{ 2 \pi n }{L} \, H \; 
\delta _{\mu 0} \, , 
\hbo 0 \le n < N_c, \; \; \; H = \mathrm{diag}(1,\ldots 1, 1-N_c) \, / N_c ,
\label{eq:3}
\ee 
where $L$ denotes the size of the torus in any direction, is still 
an invariance of the effective action. If $[P]$ denotes the 
space-time average of a particular configuration, we can map 
each configuration to a centre sector $n$ by 
\be 
c\Bigl([P]\Bigr) =n : \hbo  Re\Bigl( \tr P^\dagger \, Z_n \Bigr) 
\to \mathrm{max} , \; \; \; Z_n \; = \; \exp\{ i \, 2 \pi n \, H \} 
\label{eq:4}
\ee 
In figure~\ref{fig:1}, picture in the middle, the three centre sectors 
mark the the three global minima of the effective action. 
The pure Yang-Mills configuration space bears enough entropy that 
transitions between these centre sectors do still occur 
even at large volumes. These transitions are then associated with 
colour confinement. Only at high temperatures, the centre symmetry 
{\it spontaneously breaks}, centre sector transitions cease to exist 
and the deconfined gluon plasma phase is adopted. 

\vskip 0.3cm 
Pure Yang-Mills theories differ from QCD-like theories by the presence 
of dynamical matter which transforms under the fundamental representation 
of the group. In QCD-like theories, the $Z_N$ symmetry is 
{\it explicitly} broken by the matter-gluon interactions. 
The effective potential (see figure~\ref{fig:1} right panel) 
now possesses a unique global minimum which belongs to the 
trivial centre sector. At least for heavy matter, the degeneracy 
is only lifted by a small amount, and the questions rises whether 
transitions between centre sectors do still take place in such theories. 
A Litmus paper which is sensitive to sector transitions was 
constructed in~\cite{Langfeld:2011rh}: Dividing the spatial volume 
into two regions $V_>$ and $V_<$ of equal size, we calculate 
the spatial average of the Polyakov line, $P_>$ and $P_<$, over each 
of these regions (see figure~\ref{fig:2}). Each region of the lattice is then
mapped to  a centre sector by $c(P_{</>})$ (\ref{eq:4}). The ``Litmus paper'' is
then given by the probability $p$ that both regions belong to different
sectors.  If for an $SU(N_c)$ QCD-like theory, the centre sectors are 
attained at random in  both regions, we find $p = 1 - 1/N_c$. 
By contrast, if sector transitions have stopped at high temperatures, 
both regions belong to the {\it same} centre sector implying $p=0$. 
The result for the so-called tunneling coefficient $p$ for the pure SU(2)
gauge theory as a function of the Wilson $\beta $ coefficient is 
shown in figure~\ref{fig:2}, right panel. The finite temperature 
deconfinement phase transition is clearly visible as the transition
from $p\approx 1/2$ to $p \approx 0$. We find qualitatively the same behaviour 
for the SU(2) Higgs theory though the transition occurs at a smaller 
value of $\beta $ indicating a slip of the critical temperature 
to smaller values. Most importantly, $p \approx 1/2$ in the string-breaking 
phase at small $\beta $ indicating that centre sector transitions 
do frequently occur. 

\subsection{ \it Impact of centre-sector transitions} 

Let us now explore the phenomenological impact of the centre-sector 
transitions if dynamical quarks are included. To start with quarks 
satisfy anti-periodic boundary conditions. If $A_\mu (x)$ is 
a generic gluonic background field, we denote the quark contribution to 
the QCD partition function by the quark determinant 
$\Det _{\mathrm{AP}} [A_\mu]$ (where $AP$ is indicating the anti-periodic
boundary conditions  for the quarks). If centre-sector transitions do occur,
the gauge fields  $A_\mu ^z (x)$ in (\ref{eq:3}) are also a generic part of
the  gluonic ensemble average. Using a change of variables, the so-called 
Roberge-Weisz transformation (see~\cite{Langfeld:2009cd} for details, 
one shows that the identity for the probabilistic measure 
$$ 
\Det _{\mathrm{AP} } \Bigl[ A_\mu ^z] \; \exp \{ - S_\mathrm{YM}[A_\mu^z] \} 
\; = \; \Det _{(n) } \Bigl[ A_\mu] \; \exp \{ - S_\mathrm{YM}[A_\mu]
\}  \; ,  
$$
where the determinant on the right hand side is obtained by integrating 
quark fields which satisfy {\it $Z_n$-periodic boundary conditions}: 
$$ 
q(x + L \, \mathrm{e}_0 ) \; = \; (-1) \; Z_n \; 
q(x) \; , \hbo Z_n  \; \; \; \hbox{in (\ref{eq:4}) . } 
$$
Hence, the partition function of QCD (or the QCD-like) theory can be written 
as 
\be 
\int {\cal D} A_\mu \; \Det _{\mathrm{AP} } \Bigl[ A_\mu ] \; 
\exp \{ -S_\mathrm{YM}[A_\mu] \}  \; = \; 
\int {\cal D} A^{(0)}_\mu  \; \Biggl( \sum _n \Det _{(n) } \Bigl[ A^{(0)}_\mu] 
\, \Biggr) \; \exp \{ - S_\mathrm{YM}[A^{(0)}_\mu]
\}  \; ,  
\label{eq:5}
\ee 
where the gauge fields $A^{(0)}_\mu$ are only drawn from the trivial 
centre sector. 

\vskip 0.3cm 
The phenomenological consequences of (\ref{eq:5}) for the theory 
at {\it finite densities} was spelled out in~\cite{Langfeld:2009cd}. 
For an even number of colours $N_c$, $Z_{N_c/2}=(-1)$ is an element 
of the group implying the sum over the determinants in (\ref{eq:5}) 
includes a quark determinant where the quarks obey 
periodic boundary conditions. This implies that the quark fields 
possess Bose-statistics, and the quark operator possesses a 
mode with vanishing Matsubara frequency. If the baryon chemical 
potential approaches the quark mass gap, the $n=N_c/2$ contribution 
in $N_c$ dominates, and, at low temperatures, the quark free energy 
develops a logarithmic singularity reminiscent that of the Bose-Einstein 
condensation. This has been called {\it Fermi-Einstein condensation} (FEC). 
We stress here that for FEC to occur, 
the centre-sector transitions still need to take place since 
otherwise the sum in (\ref{eq:5}) would collapse to the contribution 
$n=N_c$ from the trivial centre sector due to the {\it spontaneous 
breakdown of centre symmetry}. Hence, FEC only applies to the confinement 
phase for which quarks cannot be considered as asymptotic states, and 
no contradiction to the spin-statistic theorem occurs. In this case, 
the observable degrees of freedom are hadrons while quarks can be viewed as 
auxiliary fields (as e.g.~ghosts fields are for the gluonic sector 
if we choose to work in a gauge fixed environment). 

\vskip 0.3cm 
FEC has been studied at great length~\cite{Langfeld:2011rh} 
in the Schwinger-model at finite densities, which can be solved 
exactly~\cite{Schwinger:1962tp,Joos:1990km,Sachs:1992pa}. 
It was found that the centre-sector transitions do take place 
even under extreme conditions and imply that the partition function 
becomes independent of the chemical potential thus solving the 
Silver-Blaze problem in this model.

\section{Cold, but dense SU(3) quark matter }

\begin{figure}[tb]
\begin{center}
\begin{minipage}[t]{8 cm} 
\centerline{ 
\epsfig{file=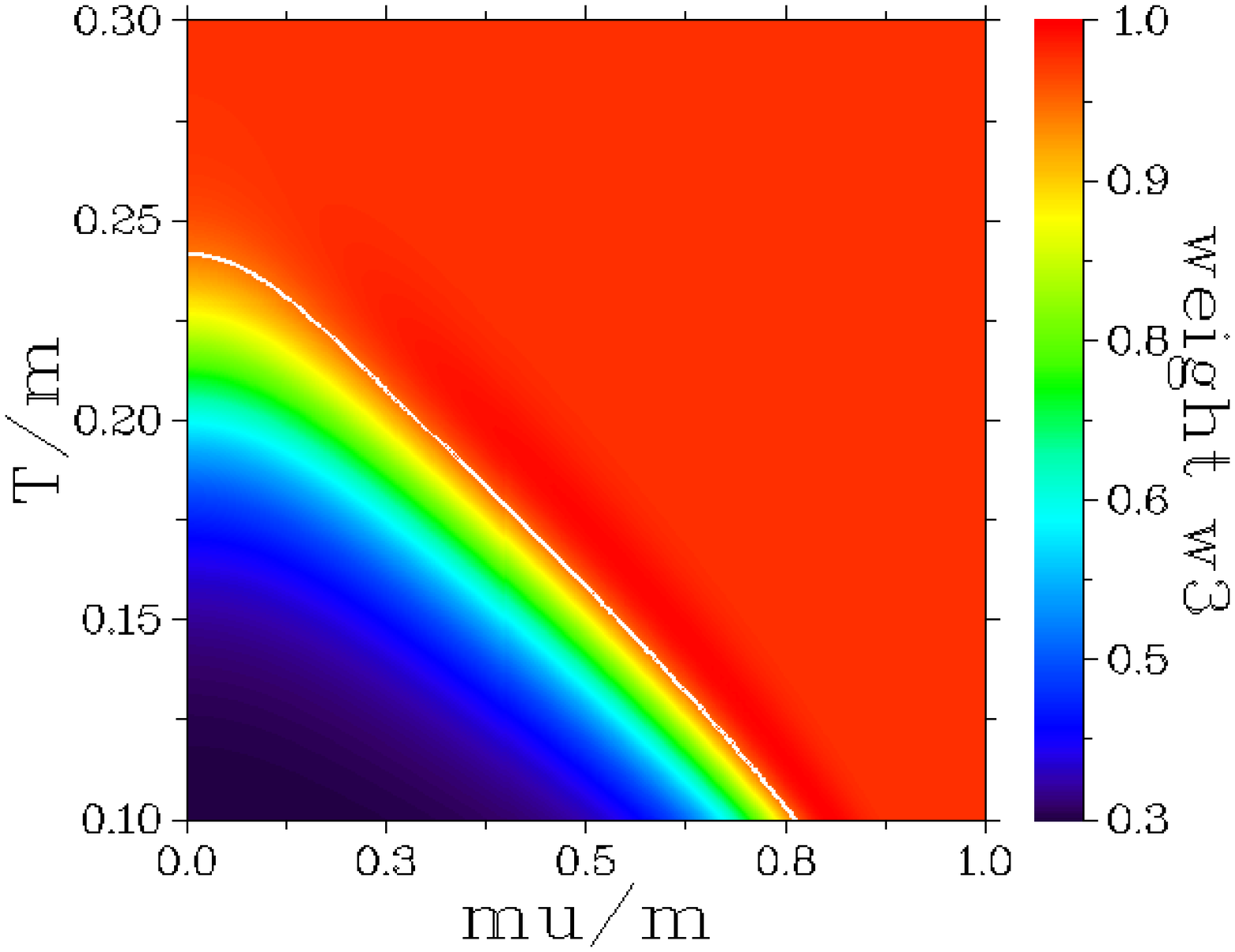,scale=0.5} \hspace{0.5cm}
\epsfig{file=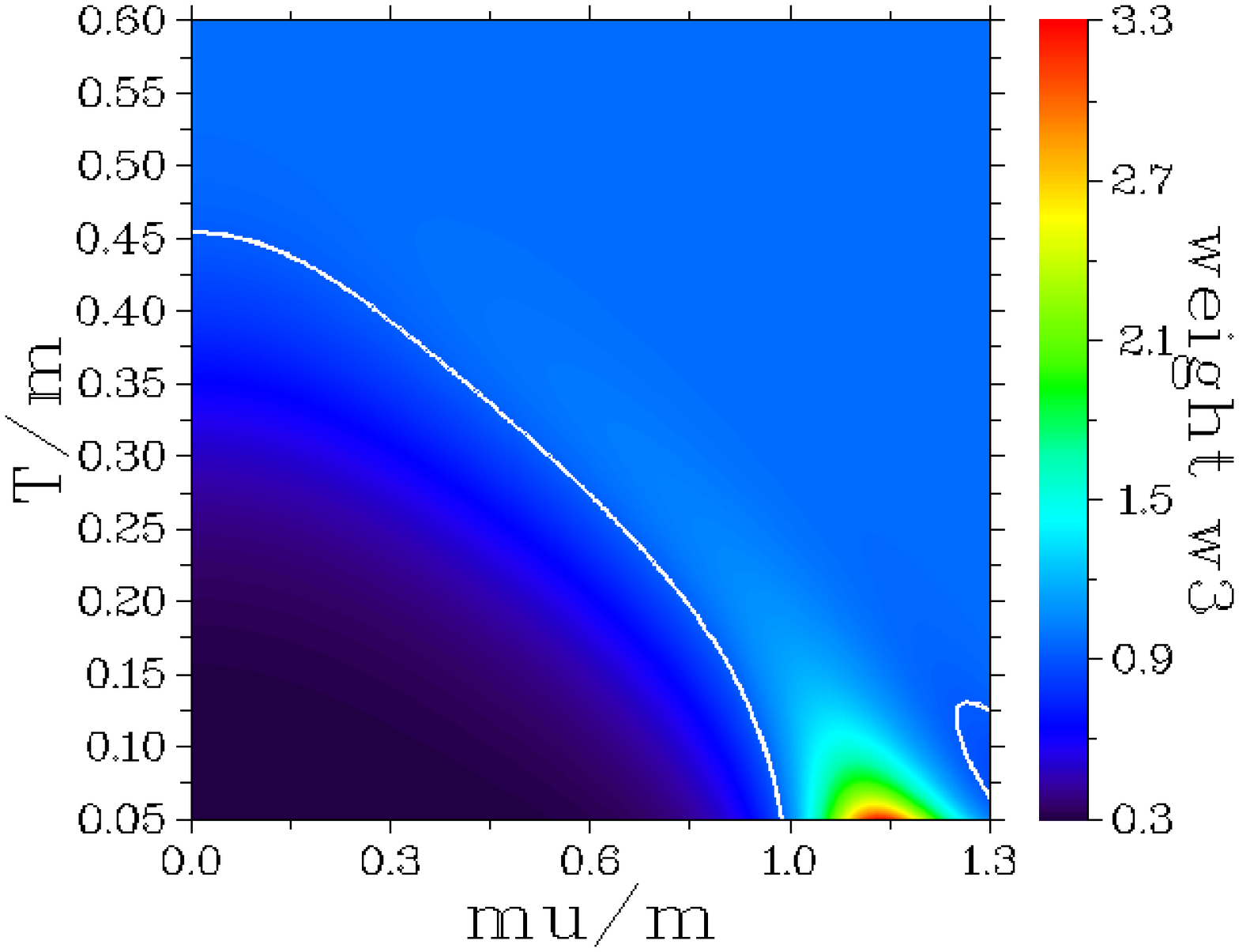,scale=0.5}
}
\end{minipage}
\begin{minipage}[t]{16.5 cm}
\caption{The phase diagram of the SU(3) quark model for a large 
($mL=15$, left panel) and a small ($mL=5$, right panel); 
note the different scale in the colour coding; 
plots from~\cite{Langfeld:2011rh}.
\label{fig:3}}
\end{minipage}
\end{center}
\end{figure}
So far, FEC is restricted to QCD-like theories with an even number of
colours. The question arises whether there is any trace of FEC left 
in theories with an {\it odd} number of colours and, most importantly, 
in QCD for $N_c=3$. Even after gauge fixing, the left hand side of
(\ref{eq:5}) would hardly admit a perturbative expansion with respect to small
gauge 
fields since centre copies  of $A_\mu=0$ are equally relevant and correspond
to large gauge fields.  On the other hand, these ``large configurations'' have
been taken into  account in the right hand side by the centre sector sum, and
an expansion with respect to $A^{(0)}_\mu$ might be perfectly viable. For the
motivation of our quark model, we take this reasoning to the extreme and set 
$A_\mu ^{(0)} =0$. This model is unrealistic since it does not produce 
a confining scale from the gluonic sector. It, however, guarantees the 
absence of coloured state from the partition function in the confinement 
phase~\cite{Langfeld:2011rh}. 

\vskip 0.3cm 
The partition function of our model is given by 
\bea
Z_Q &=&  \sum _{n=1}^{N_c} \, \int {\cal D} q{\cal D} \bar{q} \; 
\exp \left\{\bar{q} \big(\I\dslash + (A_0^{(n)} + \I\mu) \gamma^0 +  \I m
\big) q \right\}  
\; = \; \sum _{n=1}^{N_c} \E ^{\,\Gamma ^{(n)}} , 
\label{eq:eff5} \\ 
\Gamma ^{(n)} &=& \ln \,  \det \Bigl( \I\dslash + (A_0^{(n)} + \I\mu) \gamma_0 +
im  \Bigr) \; , \hbo A_0^{(n)} \; = \; 2 \pi n \, T  \, H 
\label{eq:eff6}
\eea  
where $H$ was defined in (\ref{eq:3}), $T$ is the temperature, $m$ is the
quark mass and $\mu $ is quark chemical potential. The model can solved
exactly. We here focus on the baryon number which can be written as 
\be 
Q_B \; = \;  T \; \frac{ \partial \, \ln Z_Q}{\partial \mu } 
\; = \; \sum _n \, \omega _n \; \sum _p \, \left[ \frac{ z_n^\ast }{ 
\E^{ \beta (E(p) - \mu)} + z_n^\ast  } \; - \; \frac{ z_n }{ 
\E^{ \beta (E(p) + \mu )} + z_n }  \; \right] \; , 
\label{eq:6}
\ee
where $z_n$ are the centre phases and where the centre sector weights 
$\omega _n$ are given by 
\be 
z_n \; = \; \exp\left\{ i \, \frac{2\pi}{3} n \right\} , \hbo 
\omega _n \; = \; \frac{ \exp \{ \Gamma  ^{(n)}_\mathrm{den} \} }{ 
\sum _n  \exp \{ \Gamma  ^{(n)}_\mathrm{den} \} } \; . 
\label{eq:eff25}  
\ee
The weights $\omega _n$ can be easily calculated within the model 
(see~\cite{Langfeld:2011rh} for details). 

\vskip 0.3cm 
For vanishing chemical potential $\mu =0$, 
the weights $\omega _n$ are real, positive and smaller (or equal) $1$. 
For this reason, they can be interpreted as the probability with which 
each of the centre sectors $n$ contribute to the baryon number. 
It turns out that for small temperatures ($T\ll 0.2m$), the centre 
sectors roughly contribute with equal weights, $\omega _1 \approx 
\omega _2 \approx \omega _3 \approx 1/3$, while the centre-symmetry 
also breaks spontaneously for $T \stackrel{>}{_\sim } 0.2m $. In the latter 
case, we find $\omega _1 \approx \omega _2 \approx 0$, $\omega _3 \approx 1$ 
and our model merges with the standard Fermi-gas model. 

\vskip 0.3cm 
For non-vanishing chemical potential $\mu \not=0$, the weights 
$\omega _n$ are no longer real and cannot be interpreted as sector 
probabilities anymore. Nevertheless, we still find almost an 
equal distribution for $\omega _1 \ldots \omega _3$ in the hadronic phase, 
while $\vert \omega _3 \vert \approx 1$ holds under extreme conditions. 
In this sense, we can use $\vert \omega _3 \vert $ to trace out the 
phase diagram as a function of the chemical potential $\mu $ and 
the temperature $T$. Our findings~\cite{Langfeld:2011rh} are summarised 
in figure~\ref{fig:3}. For large spatial volumes $mL>15$, 
figure~\ref{fig:3}, left panel, we find a hadronic phase separated from a
deconfinement phase under extreme conditions. 
Under pressure, i.e., for small spatial volumes $mL<5$, we find a region 
of the phase diagram at low temperatures and intermediate values for which 
$\vert \omega _3 \vert \gg 1$ holds. In this region, the quark free 
energy logarithmically diverges, and we find an excess of the baryon number if
compared to the standard Fermi gas model for same temperature and chemical
potential. From here, we conclude that, at least under pressure, 
the new state of cold, but dense matter might also be realised in 
QCD-like theories with an odd number of colours. 

\vskip 0.3cm 
Let us finally comment on the possibilities to render the above 
SU(3) quark model more realistic: (i) In a phenomenological oriented 
approach, one might consider an NJL-type four fermion interaction besides of 
the average over the centre-sector background fields. Here, it would be 
interesting to see whether a spontaneous breakdown of chiral symmetry 
also implies a spontaneously broken centre symmetry and hence 
deconfinement. (ii) In a more fundamentally oriented approach, one notices 
that the prototypes of centre sector gauge fields configurations 
are homogeneously stretching throughout space-time. This is a state of low 
energies but with little entropy. Breaking up the gauge fields into 
patches of different centre-sector fields significantly increases 
the entropy and costs energy proportional to the interfaces between 
the patches. The interfaces can be identified with centre-vortices, and 
it was observed e.g.~in~\cite{Kovacs:2000sy} that the vortex interface energy 
vanishes in the confinement phase leading to percolating vortices. 
Here, the so-called planar vortex density sets the confinement 
scale~\cite{Langfeld:1997jx}. Hence, supplementing quarks to a SU(3) vortex
model would be natural extention of the SU(3) Fermi gas model. 

\vskip 0.3cm 
\noindent
{\bf Acknowledgments: } 
This work is in parts a project of the DiRAC framework, supported 
by STFC.


\begin{thebibliography}{99}
\itemsep -2pt 

\bibitem{Svetitsky:1982gs}
  B.~Svetitsky, L.~G.~Yaffe,
  \Journal{\NPB} {210} {423} {1982}

\bibitem{Svetitsky:1985ye}
  B.~Svetitsky,
  \Journal{\em Phys. Rept.} {132} {1-53} {1986} 

\bibitem{Greensite:2003bk}
  J.~Greensite,
 \Journal{\em Prog. Part.Nucl.Phys.} {51} {1} {2003} 

\bibitem{Zakharov:2006vt}
  V.~I.~Zakharov, {``Matter of resolution: From quasiclassics to fine
    tuning''},  {\em [hep-ph/0602141]} 

\bibitem{Langfeld:1997jx}
  K.~Langfeld, H.~Reinhardt, O.~Tennert,
  \Journal{\PLB} {419} {317} {1998}

\bibitem{Langfeld:2010nm}
  K.~Langfeld, E.~-M.~Ilgenfritz,
  \Journal{\NPB} {848} {33-61} {2011} 


\bibitem{Langfeld:1998cz}
  K.~Langfeld, O.~Tennert, M.~Engelhardt, H.~Reinhardt,
  \Journal{\PLB} {452} {301} {1999} 

\bibitem{Engelhardt:1999fd}
  M.~Engelhardt, K.~Langfeld, H.~Reinhardt, O.~Tennert,
  \Journal{\PRD} {61} {054504} {2000}

\bibitem{Langfeld:2003zi}
  K.~Langfeld,
  \Journal{\PRD} {67} {111501} {2003}

\bibitem{Langfeld:2009cd}
  K.~Langfeld, B.~H.~Wellegehausen and A.~Wipf,
  \Journal{\PRD} {81} {114502} {2010}

\bibitem{Langfeldect2010}
K.~Langfeld, {``Centre-sector tunneling, confinement and
the quark Fermi surface''}, 
invited talk at the workshop on "Chiral symmetry and confinement in cold 
dense quark matter", ECT, Trento, July 19 - 23, 2010. 

\bibitem{Langfeld:2011rh}
  K.~Langfeld, A.~Wipf,
  {``Fermi-Einstein condensation in dense QCD-like theories''}, 
  {\em [arXiv:1109.0502 [hep-lat]]}

\bibitem{Keurentjes:1998uu}
  A.~Keurentjes, A.~Rosly and A.~V.~Smilga,
  \Journal{\PRD} {58} {081701} {1998} 

\bibitem{Selivanov:1999cr}
  K.~G.~Selivanov,
  \Journal{\PLB} {471} {171} {1999} 

\bibitem{Schaden:2004ah}
  M.~Schaden,
  \Journal{\PRD} {71} {105012} {2005} 

\bibitem{Schwinger:1962tp}
  J.~S.~Schwinger,
  \Journal{\PREV} {128} {2425-2429} {1962}

\bibitem{Joos:1990km}
  H. Joos,
  \Journal{\em Helv. Phys. Acta } {63} {670-682} {1990}

\bibitem{Sachs:1992pa}
I.~Sachs and A.~Wipf,
  \Journal{\em Helv. Phys. Acta } {65} {652-678} {1992}

\bibitem{Kovacs:2000sy}
  T.~G.~Kovacs, E.~T.~Tomboulis,
  \Journal{\PRL} {85} {704-707} {2000} 


\end{thebibliography}
\end{document}